\documentclass[11pt]{article}
\usepackage{moriond,epsfig}

\def\be{\begin{equation}}
\def\ee{\end{equation}}
\def\bea{\begin{eqnarray}}
\def\eea{\end{eqnarray}}

\newcommand{\met}       {\mbox{$\not\!\!E_T$}}

\newcommand{\ttbar}     {\mbox{$t\bar{t}$}}
\newcommand{\qqbar}     {\mbox{$q\bar{q}$}}
\newcommand{\ppbar}     {\mbox{$p\bar{p}$}}
\begin{document}
\vspace*{4cm}
\title{MEASUREMENT OF THE TOP QUARK AND W BOSON MASSES}

\author{ R. L. KEHOE (for the D0 and CDF Collaborations)}

\address{Department of Physics, Southern Methodist University,\\
Dallas, Texas 75275}

\maketitle\abstracts{
We measure the top quark mass in \ttbar\ events using up to 3.6~fb$^{-1}$ 
of \ppbar\ collisions at $\sqrt{s}=2$~TeV.  New results are described
for matrix element methods with single lepton and dilepton channels,
and template methods with all-jets and dilepton channels.
Standardization of systematic uncertainties has proceeded between
the CDF and D0 collaborations. A combined measurement of 
$173.1\pm0.6(\rm stat)\pm1.1(syst)$~GeV is achieved. Fits to all measured electroweak 
parameters gives a Higgs boson mass 95\% CL upper limit of $m_H < 163$~GeV.
We also present a new measurement of the $W$ boson mass from D0. Three
different template methods are performed. Accounting for their correlations,
a combined measurement of $M_W=80.401\pm0.043(\rm stat+syst)$~GeV is obtained.}

\section{Introduction}

The study of the top quark and $W$ boson masses continues to be important in electroweak
physics.  Radiative corrections to the calculation
of the $W$ boson mass, $M_W$, include terms which are logarithmic in the Higgs
mass, $m_H$, and vary as the square of the top quark mass, $m_t$.  This allows
a constraint on $m_H$ from measurements of the other two parameters.
Another source of
interest in the top quark mass has been the peculiar unitary value
of the corresponding Yukawa coupling, $Y_t$. As a result, the study of both the $W$ boson
and top quark masses are major goals of the Fermilab Tevatron program.  I present
recent measurements of both quantities in this paper.

Experimentally, the measurements are currently made exclusively at the Fermilab Tevatron
by the D0 and CDF experiments.  These are large, multipurpose collider detectors capable
of low background particle identification for electrons and muons, and good momentum measurements
for leptons, jets and \met.  Recorded data samples amount to 6 fb$^{-1}$ per experiment.  
The measurement of $m_t$ has been undergoing
rapid improvements in precision over the last few years. Up to
3.6~fb$^{-1}$ have been analyzed so far, which represents a ten-fold increase from data
available just four years ago. Improved methods have permitted more precise 
measurements.  Advances in the estimation of systematic effects have substantially
reduced remaining uncertainties.

\section{Methods of Top Quark Mass Measurement}

In general, the methods used to measure the top quark mass fall into
two categories (for exception, see Ref.~\cite{sharyy}).  Template methods fit a
set of quantities, $x_i$, derived from event observables such as the measured
kinematic quantities.  The $x_i$ are correlated with $m_t$ and are compared to
distributions expected from top quarks of varying mass.
Often, template methods involve a kinematic reconstruction of the
event by solving constraint equations, and then fitting to the solved
top quark mass. Matrix element methods instead compare the observables, 
the $x_i$ of this method, directly to expectations using a leading
order (LO) matrix element calculation for \ttbar. In this way, event-by-event
likelihoods are generated which use all of the kinematic information
in an event. In each of these approaches, the background is accounted for
by modeling the $x_i$.  A probability of consistency
vs. $m_t$ is then achieved by accounting for \ttbar\ and background:
\begin{equation}
P(x_{i}|m_{t})=f_{top}P_{top}(x_{i}|m_{t})+f_{bkg}P_{bkg}\left(x_{i}\right)
\label{eq:prob}
\end{equation}
where $P_{top}$ and 
$P_{bkg}$ are signal and background probability densities based on $x_i$ and a
particular $m_t$. For matrix element approaches, $P_{top}$ incorporates
the \ttbar\ matrix element.  For a particular final state channel:
\begin{equation}
P_{top}(x_i |m_{t})=\frac{1}{\sigma_{obs}(\qqbar\to\ttbar\to {\rm channel} |m_{t})}\times\int\sum dq_{1}dq_{2}f(q_{1})f(q_{2})\frac{(2\pi)|M|^{2}}{4\sqrt{(q_{1}\cdot q_{2})^{2}}}\cdot d\Phi_{6}\cdot W(x_i,y)
\label{eq:melem}
\end{equation}
where $M$ is the LO matrix element, and $f(q)$ are the parton distribution functions.

The precision of the world average measurement of $m_t$ has become 
dominated by systematic uncertainties.
The calibration of jet energies particularly limits the achievable precision of the
$m_t$ measurement. However,
top quark events offer an $in$ $situ$ dijet resonance in channels where 
$W\to q\bar{q}'\to jj$.  As a result, measurements in
these channels typically perform a simultaneous fit to the jet energy
scale (through $m_{jj}$) and to $m_t$. This allows the top quark mass to be evaluated at the scale
favored by the correct $M_W$, and the residual systematic uncertainty
is reduced.  Systematic uncertainties have been defined to
facilitate measurement combinations.
Dominant physics modeling uncertainties come from hadronization, underlying 
event, and the modeling of backgrounds. Major
detector modeling uncertainties arise from the residual jet energy
scale, and also the difference between $b$-jet and
light-quark jet calibrations. 
Template methods can have significant uncertainty due to the statistical 
samples available for template generation.

\section{All-hadronic Channel}

In a \ttbar\ event, both quarks are expected to decay to a $W$ boson
and a $b$ quark. An all-hadronic ($6j$) final state results when both $W$ bosons
decay to quark pairs. Such a final state has an enormous background
from QCD multijet production.  This is ameliorated with the use of
multiparameter discriminants and the tagging of $b$-jets by identification
of their displaced decay vertices from long $b$ hadron lifetimes. CDF
has performed a mass analysis in 2.9 fb$^{-1}$ of collider data using this
channel~\cite{cdf6j}. To enhance their sensitivity, they have developed a further
discriminant which separates quark jets from gluon jets.  The resulting
background reduction gives a 25\% improvement in statistical uncertainty
from the mass measurement.

\begin{figure}
\psfig{figure=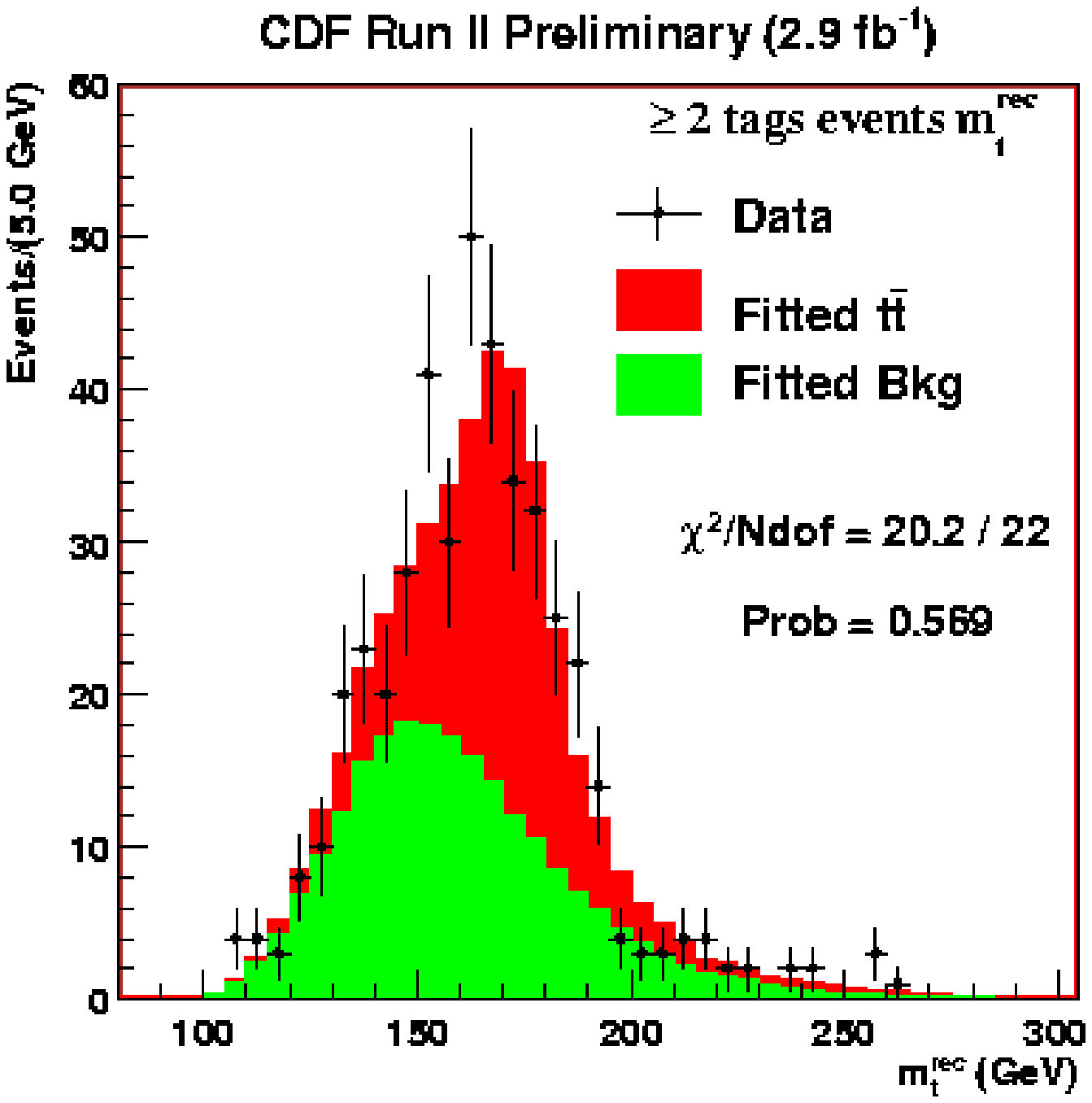,height=3.0in}
\psfig{figure=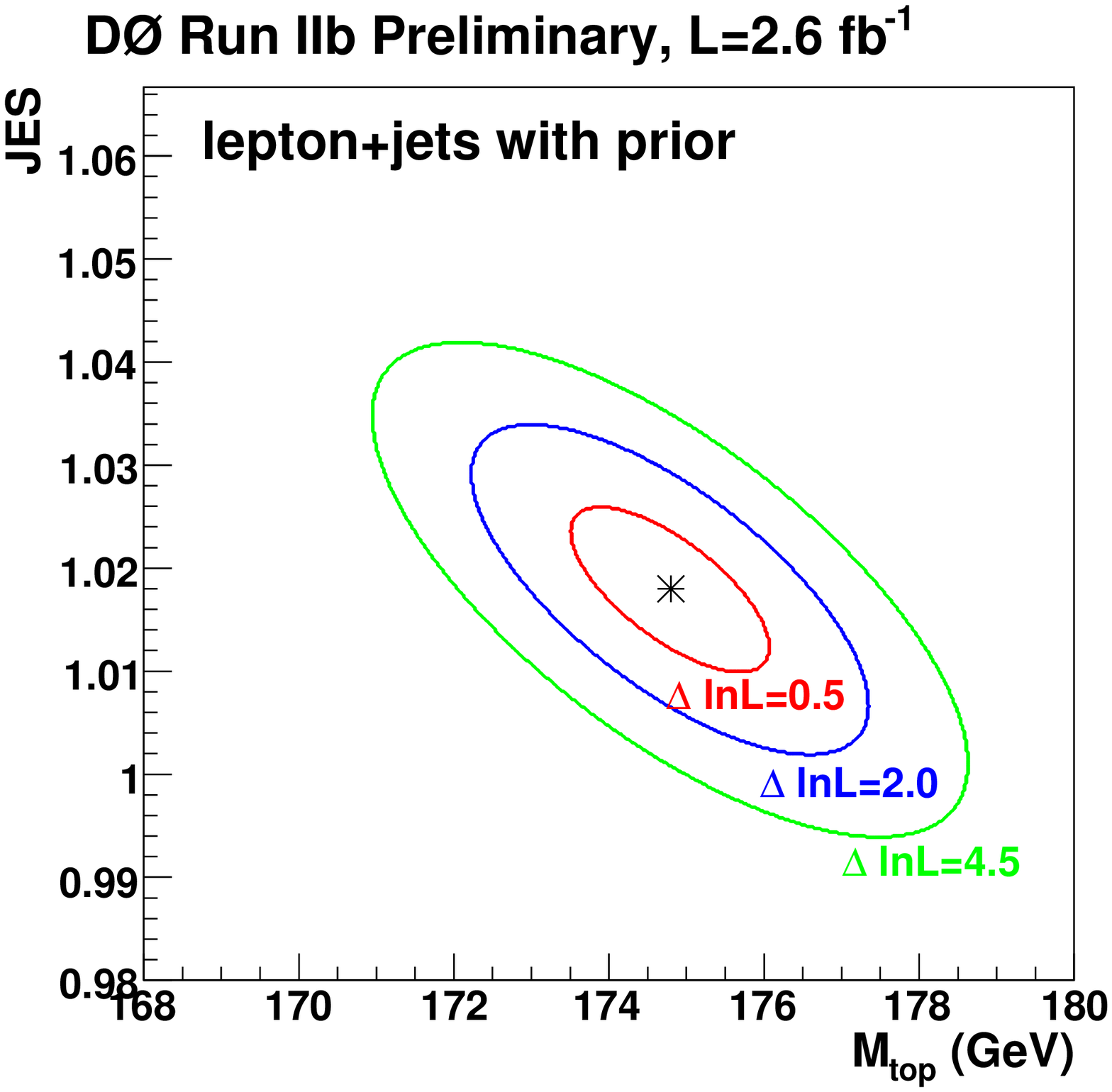,height=3.0in}
\caption{Left: CDF reconstructed $m_t$ from $6j$ events.  Fitted \ttbar\ and
background contributions are overlaid with data.  Right: D0 fitted
jet energy scale vs. $m_t$ for $\ell+$jets events.  Contours of constant
$\Delta \mathcal{L}$ are shown.
\label{fig:mtplots}}
\end{figure}

The method of mass measurement proceeds by calculating dijet and three
jet masses for each of the different assignments of jets to $W$ bosons and
$b$ quarks. A $\chi^2$ is calculated of these masses with respect to fitted
top quark and $W$ boson masses, as well as fitted momenta, given expected widths
and resolutions of these parameters. The shapes of the reconstructed
$W$ boson and top quark masses are obtained by fitting shapes from the \ttbar\ and
background samples. These are used as templates to compare to data.
The result of this fit is shown in Fig.~\ref{fig:mtplots}.
The use of the $M_W$ constraint reduces the systematic uncertainty,
but the residual jet energy scale is still a large contributor, along
with color reconnection and residual bias errors. 

\section{Single Lepton Plus Jets Channels}

Final states which provide high statistics but also reasonable
background levels are those where at least one $W$ boson from top quark
decay itself decays to a lepton-neutrino ($\ell\nu$, where $\ell=e,\mu$) 
pair. These $\ell+$jets events have been updated recently by both Tevatron 
experiments using matrix element approaches.  D0 has selected events in
2.6 fb$^{-1}$ of Run 2b data using a neural network $b$-tag selection~\cite{d02bMElj}.
For both the signal and background probabilities given in Eq.~\ref{eq:prob}, matrix
elements are used. 
As shown in Fig.~\ref{fig:mtplots}, the fitted jet energy scale is $1.018\pm0.007$, which is
consistent with the jet energy scale uncertainty from $\gamma+$jet events.
The error is also
within the expectation of pseudoexperiment tests from the Monte Carlo.
The dominant systematic uncertainty comes from the relative $b$/light
jet response uncertainty of 0.8~GeV.

Using 3.2 fb$^{-1}$, CDF has selected events with an exclusive four-jet 
selection~\cite{cdf2MElj}.
One of these jets must satisfy a $b$-tag requirement. A neural network
selection is performed based on 19 kinematic quantities using a quasi-Monte
Carlo method to obtain uniformity in the integration. Not only is
a clear separation between signal and background achieved, but the
NN output is independent of top quark mass for the signal. In order
to improve the mass resolution, the likelihood which comes from the
mass fit is used to reject background and poorly reconstructed
\ttbar\ events. This is done by cutting out low peak likelihood events.
The dominant uncertainty comes from the jet energy calibration.

\section{Dilepton Channels}

The rarest events occur when both $W$ bosons decay to $\ell\nu$.  
Such dilepton
($2\ell$) channels can usually attain low backgrounds without resort
to $b$-tagging. However, there is no $W\to jj$ resonance with which to obtain
an $in$ $situ$ calibration and reduce the jet energy scale uncertainty.
In addition, there is an ambiguity of how the measured event \met\
corresponds to the two neutrino momenta. The dilepton analyses have
been pursued in both template and matrix element approaches recently.
Using $e\mu$ events in 3.6 fb$^{-1}$ of data, D0 has employed matrix
elements for both signal and background in the likelihood calculation~\cite{d0emuME}.
For the latter, the largest $Z\to \tau\tau$ background is assumed for all
backgrounds. This yields a measurement of
$m_t=174\pm3.3(\rm stat)\pm2.6(syst)$~GeV.
The main systematic uncertainties are the jet energy
scale uncertainty, and the difference between $b$-jet and light quark
jet energy scales. Hadronization and background modeling also play
significant roles.

Template methods have also been recently finalized by D0 for the three standard
dilepton channels ($ee$, $e\mu$, $\mu\mu$) plus two $b$-tagged channels with
loosened lepton identification requirements ($e/\mu+$track)~\cite{d02lPRD}. Because
the dilepton event is underconstrained due to the two neutrinos, the
kinematic reconstruction supplies an input parameter. They sample assumed
neutrino rapidity distributions to provide a relative weight vs. $m_t$. 
D0 uses the moments of the weight distribution in 1 fb$^{-1}$ to extract 
$m_t=176.2\pm4.8(\rm stat)\pm2.1(syst)$~GeV. 
D0 has also applied a template method with
the same sample using partial production and decay information.  These two
measurements combined give $m_t=174.8\pm4.4(\rm stat)\pm2.0(syst)$ GeV.

\begin{table}[t]
\caption{Recent $m_t$ measurements by CDF and D0 experiments.  The D0 $2\ell$
entry combines the $e\mu$ matrix element and $ee/\mu\mu/\ell+$track template measurements.  
The last two entries give the combined result for each experiment for Run 1 plus Run 2 data.
\label{tab:mtvals}}
\vspace{0.4cm}
\begin{center}
\begin{tabular}{|c|c|c|c|l|}
\hline
Channel & Experiment & Luminosity & $m_t$ (GeV)\tabularnewline
\hline
\hline 
$6j$ & CDF & 2.9 fb$^{-1}$ & $174.8\pm2.4(\rm stat+JES)^{+1.2}_{-1.0}(syst)$~\cite{cdf6j}\tabularnewline
\hline 
$\ell+$jets & D0 & 3.6 fb$^{-1}$ & $173.7\pm0.8(\rm stat)\pm1.6(syst)$~\cite{d02bMElj}\tabularnewline
\hline 
$\ell+$jets & CDF & 3.2 fb$^{-1}$ & $172.1\pm0.9(\rm stat)\pm0.7(JES)\pm1.1 (syst)$~\cite{cdf2MElj}\tabularnewline
\hline 
$2\ell$ & D0 & 1-3.6 fb$^{-1}$ & $174.7\pm2.9(\rm stat)\pm2.4(syst)$~\cite{d0emuME,d02lPRD}\tabularnewline
\hline 
all & CDF & up to 3.2 fb$^{-1}$ & $172.6\pm0.9(\rm stat)\pm1.2(syst)$~\cite{cdfmtop}\tabularnewline
\hline 
all & D0 & up to 3.6 fb$^{-1}$ & $174.2\pm0.9(\rm stat)\pm1.5(syst)$~\cite{d0mtop}\tabularnewline
\hline
\end{tabular}
\end{center}
\end{table}

\section{World Average and Electroweak Fits}

\begin{figure}
\psfig{figure=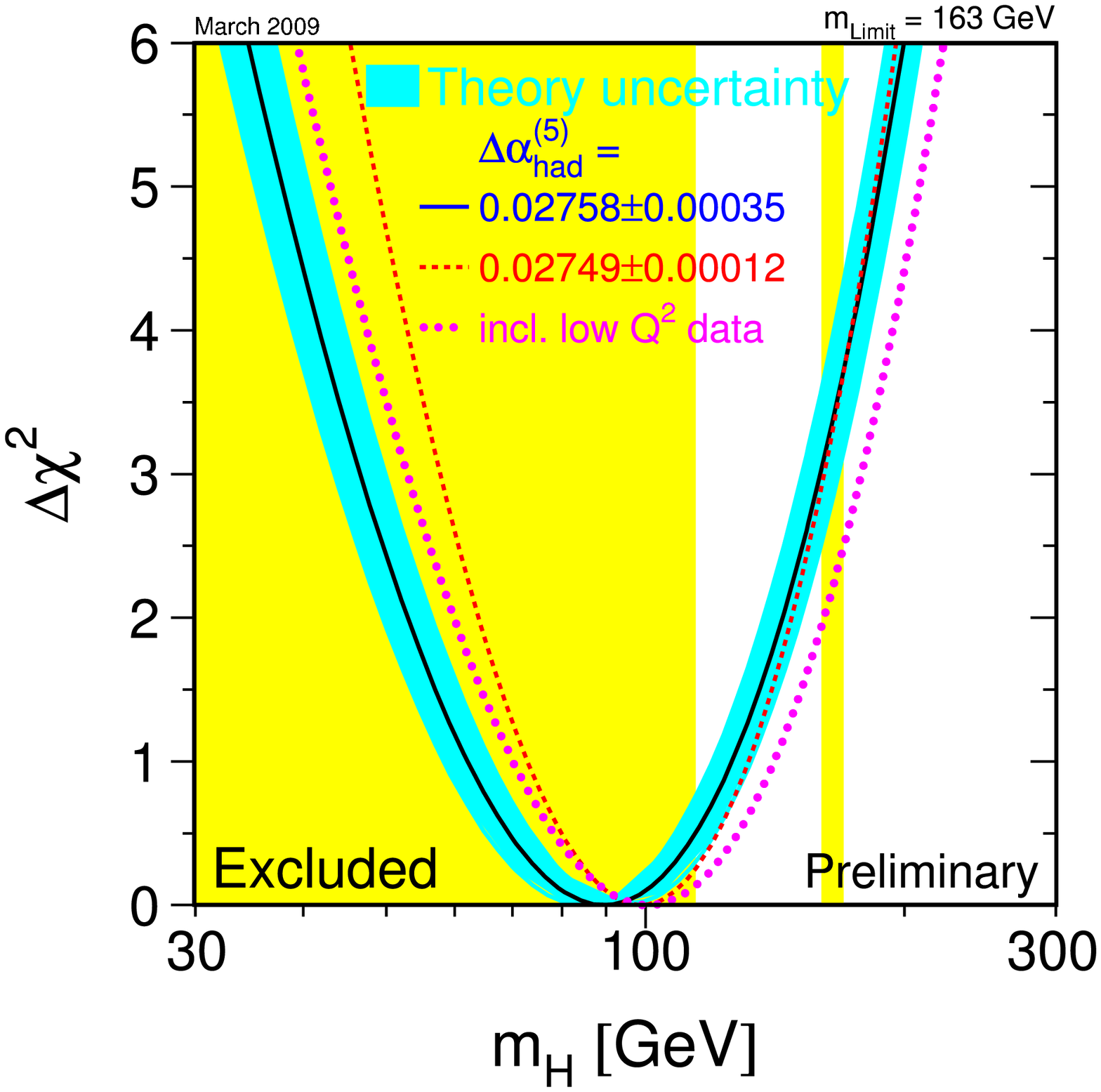,height=2.7in}
\psfig{figure=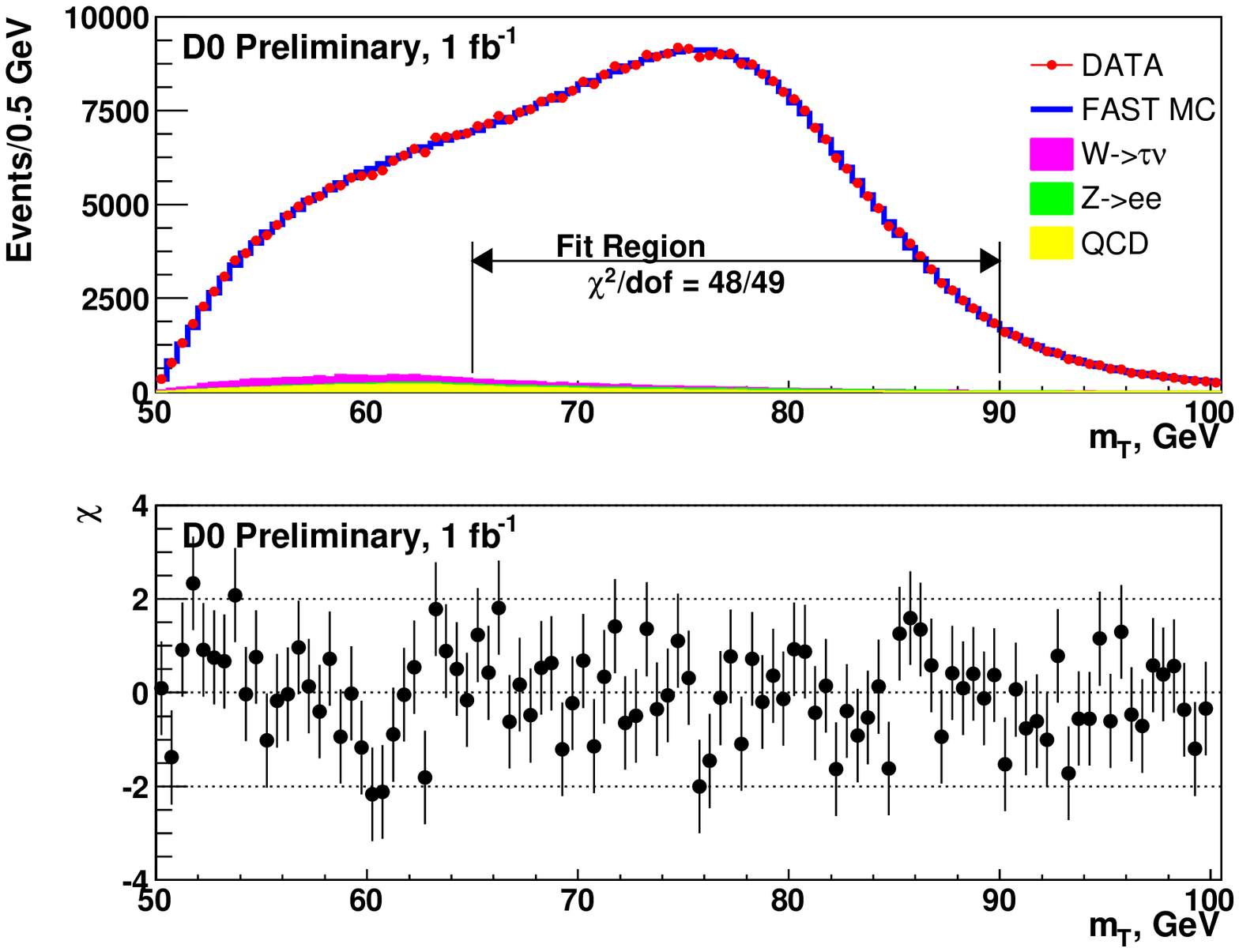,height=2.7in}
\caption{Left: Fit to $m_H$ using electroweak constraints including the new
world average $m_t$ measurement.  The $m_H<163$ GeV limit, as 
well as the 
direct 95\% c.l. exclusion band of $160-170$~GeV masses from Tevatron searches 
are indicated.  Right:
The fitted distribution of $m_T$ in $W\to e\nu$ events in 1 fb$^{-1}$ of D0 data.  
Residuals of the data relative to the fitted $W$ boson template are shown at 
lower right.
\label{fig:smfits}}
\end{figure}

The measured value of $m_t$ for $6j$, $\ell+$jets and $2\ell$ channels is given
in Table~\ref{tab:mtvals}, incorporating combinations of Run 2a and 2b results.  
Combined measurements for each experiment using all Run 1 and Run 2 results are
also indicated.  The world average top mass determined 
from these is $173.1\pm0.6(\rm stat)\pm1.1(syst)$~GeV~\cite{tevmtop}, 
which corresponds to $Y_t = 0.995\pm0.007$.  In the context of the electroweak model,
fitting all measured electroweak parameters including $m_t$ yields a
new constraint of the Higgs boson mass of $m_H<163$~GeV~\cite{ewkfit} at 95\% CL. The most likely
value is $m_H = 90^{+36}_{-27}$~GeV (see Fig.~\ref{fig:smfits}), to be compared with the 
direct LEP2 lower limit of 114.4~GeV.

\section{Measurement of the W Boson Mass}

The D0 experiment has completed a preliminary, precision
measurement of the $W$ boson mass. They use 499,830 $W\to e\nu$ events obtained
from 1 fb$^{-1}$ of collider data to perform three template analyses in
parallel.  These use the electron $p_T$, the \met\ and the transverse mass
calculated from these two, $m_{T}=\sqrt{2p_{T}^{e}p_{T}^{\nu}(1-\cos(\phi_{e}-\phi_{\nu}))}$.
This measurement requires a precise calibration of the electromagnetic
calorimeter, which is provided by the 18,725 $Z\to ee$ events in this
data sample. A detailed modeling of the 
detector response to the recoiling system is also needed. 

The analysis proceeds by comparing the distribution of the template variables
with high statistics distributions from models of signal and background.
For signal, these are obtained in 10 MeV steps using RESBOS~\cite{resbos} 
and PHOTOS~\cite{photos}. 
These templates must have $10^8$
events each, so a tuned fast simulation of the detector is used. The
$Z\to ee$ process is modeled in a similar fashion to facilitate a carryover
of the electromagnetic calibration to the $W$ events. A parametrized
functional form for the $Z$ boson reconstructed mass distribution 
is taken from GEANTed Monte Carlo
samples and tuned to the data. The fit of the data to the
templates in the mass range of 70 to 110 GeV yields the electron energy
calibration scale and offset with $\chi^2/dof = 153/160$. $Z$ boson events are
also used to study the hadronic recoil to the $W$ boson.  A GEANTed Monte
Carlo sample again provides the functional 
form for the detector response to the hard component.  Contributions due
to the spectator quarks and additional \ppbar\ collisions are modeled by fitting to
$Z$ boson collider data.  The backgrounds in this channel
arise from $Z\to ee$ events where one electron is not identified, QCD
instrumental background, and $W\to \tau\nu\to e\nu\nu$. These are modeled using
an electron plus track control sample, a trigger sample with track requirements
omitted, and a full GEANT Monte Carlo simulation, respectively.

The $W$ boson mass is measured by blinding the data sample with an unknown
offset for all three measurements simultaneously.  The blinding was done to avoid
bias by knowledge of the current world average value.  The comparison
of the data to the templates is performed, yielding good $\chi^2$ for
each measurement (e.g. $m_T$ yielded $\chi^2/dof=48/49$). Once the blinding
was removed, a measurement was obtained for each method, as given
in Table~\ref{tab:smvals} and Fig.~\ref{fig:smfits}.
The systematic uncertainties are dominated by a 34~MeV
uncertainty due to the electron energy scale. The methods are somewhat decorrelated, 
and a combination yields 
$80.401\pm0.021(\rm stat)\pm0.038(syst)$ GeV $=80.401\pm0.043$~GeV~\cite{d0wmass}. 
This will be propagated into the world average $W$ boson mass and electroweak
fits soon.

\begin{table}[t]
\caption{Measurements of $M_W$ using three different template variables.  
$W\to e\nu$ events in 1 
fb$^{=1}$ of D0 data were used.\label{tab:smvals}}
\vspace{0.4cm}
\begin{center}
\begin{tabular}{|c|c|l|}
\hline 
template variable & $M_W$ (GeV)\tabularnewline
\hline
\hline 
$m_T$ & $80.401\pm0.023(\rm stat)\pm0.037(syst)$\tabularnewline
\hline 
\met\ & $80.400\pm0.027(\rm stat)\pm0.040(syst)$\tabularnewline
\hline 
$p_T^e$ & $80.402\pm0.023(\rm stat)\pm0.044(syst)$\tabularnewline
\hline
\end{tabular}
\end{center}
\end{table}

\section{Conclusions}

Tevatron experiments have analyzed up to 3.6 fb$^{-1}$ of collider data for
top quark and $W$ boson mass
measurements. For the former, the all-jets channels are achieving
1.5\% precision, the single leptons have surpassed the 1\% precision,
and even the rare dilepton events are providing 2\% precision to the
top mass measurement. The world average is now $173.1\pm0.6(\rm stat)\pm1.1(syst)$~GeV,
which gives 0.7\% precision. This update generates a new 95\% CL
upper limit on the Higgs boson mass of 163 GeV. The $W$ boson mass has
also been measured by D0 in the $e\nu$ channel, yielding
the world's best measurement by a single experiment: $M_W=80.401\pm0.043(\rm stat+syst)$~GeV.

\section*{Acknowledgments}

We thank the staffs at Fermilab and collaborating institutions, 
and acknowledge support from the 
DOE and NSF (USA);
CEA and CNRS/IN2P3 (France);
FASI, Rosatom and RFBR (Russia);
CNPq, FAPERJ, FAPESP and FUNDUNESP (Brazil);
DAE and DST (India);
Colciencias (Colombia);
CONACyT (Mexico);
KRF and KOSEF (Korea);
CONICET and UBACyT (Argentina);
FOM (The Netherlands);
STFC and the Royal Society (United Kingdom);
MSMT and GACR (Czech Republic);
CRC Program, CFI, NSERC and WestGrid Project (Canada);
BMBF and DFG (Germany);
SFI (Ireland);
The Swedish Research Council (Sweden);
CAS and CNSF (China);
and the
Alexander von Humboldt Foundation (Germany).

\end{document}